# Social media use among American Indians in South Dakota: Preferences and perceptions

**Deepthi Kolady[1], Amrit Dumre[2], Weiwei Zhang[3], Kaiqun Fu[4], and Marcia O'Leary[5], Laura Rose[6]**

[1] Associate Professor, Ness School of Management and Economics

South Dakota State University

Deepthi.kolady@sdstate.edu

[2] Graduate Research Assistant, Ness School of Management and Economics

South Dakota State University

Amrit.Dumre@jacks.sdstate.edu

[3] Associate Professor, School of Psychology, Sociology, and Rural Studies

South Dakota State University

Weiwei.Zhang@sdstate.edu

[4] Assistant Professor, Department of Electrical Engineering and Computer Science

South Dakota State University

kaiqun.fu@sdstate.edu

[5] Director, Missouri Breaks Industries Research Inc

marcia.oleary@mbiri.com

[6] Business Manager, Missouri Breaks Industries Research Inc

laura.rose@mbiri.com

**Abstract**

Social media use data is widely being used in health, psychology, and marketing research to analyze human behavior. However, we have very limited knowledge on social media use among American Indians. In this context, this study was designed to assess preferences and perceptions of social media use among American Indians during COVID-19. We collected data from American Indians in South Dakota using online survey. Results show that Facebook, YouTube, TikTok, Instagram and Snapchat are the most preferred social media platforms. Most of the participants reported that the use of social media increased tremendously during COVID-19 and had perceptions of more negative effects than positive effects. Hate/harassment/extremism, misinformation/made up news, and people getting one point of view were the top reasons for negative effects.

## 1. Introduction

Social media platforms (SMPs) have been used as media for people to express themselves when dealing with catastrophic events such as natural disasters (Sakaki, Okazaki, and Matsuo 2010; Yates and Paquette 2011), and epidemic (Zhao, Chen, et al. 2020). There is an emerging interest in integrating social media data into health research, ranging from assessing national and local influenza infection rates (Broniatowski, Paul, and Dredze 2013; Bodnar and Salathé 2013), modeling human behaviors and social networks related to certain health outcomes (Gore, Diallo, and Padilla 2015; Young, Rivers, and Lewis 2014), and predicting the level of stress and anxiety and the onset of depression (Stieglitz and Dang-Xuan 2013; De Choudhury et al. 2013; Liu et al. 2017). Studies using data collected for various SMPs discovered that during the COVID-19 outbreaks, the discussions contained positive and negative sentiments and temporal trend in positive and negative sentiments may correspond to the direct impact of the pandemic, such as COVID-19 mortality, new variants, vaccinations, and its subsequent impacts on the socio-economic environments (Li et al. 2020; Zhao, Cheng, et al. 2020; Das and Dutta 2021).

However our understanding of social media use among American Indians, one of the minority groups adversely affected by COVID-19, is limited (Zhang and Kolady 2022; Kolady and Zhang 2023). The goal of this study is to understand American Indian's preference for social media use in general and particularly during COVID-19.

## 2. Objectives

The broad objective of the study was to investigate American Indian's preferences and perceptions of social media type and examine how social media was used by American Indians during COVID-19. The specific objectives of the study were to

- Identify use preference for social media type,
- Examine the impact of COVID-19 on social media use,
- Assess the perceived effects of social media use.

## 3. Method

Structured survey questionnaire was prepared and finalized in February 2023. An online survey was employed from March to April 2023 for collecting data from American Indians after getting the Institutional Review Board approval from South Dakota State University. The research team worked with Missouri Breaks Industries Research Inc to circulate the online survey among American Indians in South Dakota. We used identity as American Indian, age above 18 years old, and respondents belonging to South Dakota as screening variables. We only obtained 25 responses meeting all these criteria. We then analyzed the collected data to assess the extent of social media use among American Indians in South Dakota.

Summary statistics of results in terms of frequency and percentage were calculated for most of the variables. Reasons cited for positive and negative effects of social media use were ranked in accordance with the weight given to each reason (7=most important to 1= least important) (See Figure 5). It was calculated by using formula.

$$Total\ score = \sum_{i}^{n} S_i F_i$$

Where, i= Each individual response

S= Scale (weight) given to the option

F = Frequency of scale (weight)

For example, total score calculated for reasons of positive effect of social media use was highest (42) for "Helping people informed and aware". It was calculated as;

$$Total\ score = 5\ X\ 7 + 1\ X\ 4 + 1\ X\ 3 = 42$$

## 4. Key findings
### 4.1 Sociodemographic characteristics

Out of the 25 respondents, majority (72%) of them were male. With respect to the age group, most of the respondents (40%) were of the age group 55 to 64. Most of the respondents (68%) had some college degree. Most of the respondents (76%) lived on reservations. Further, 68% of the respondents lived about 75-100% of the time on reservations during the COVID-19.

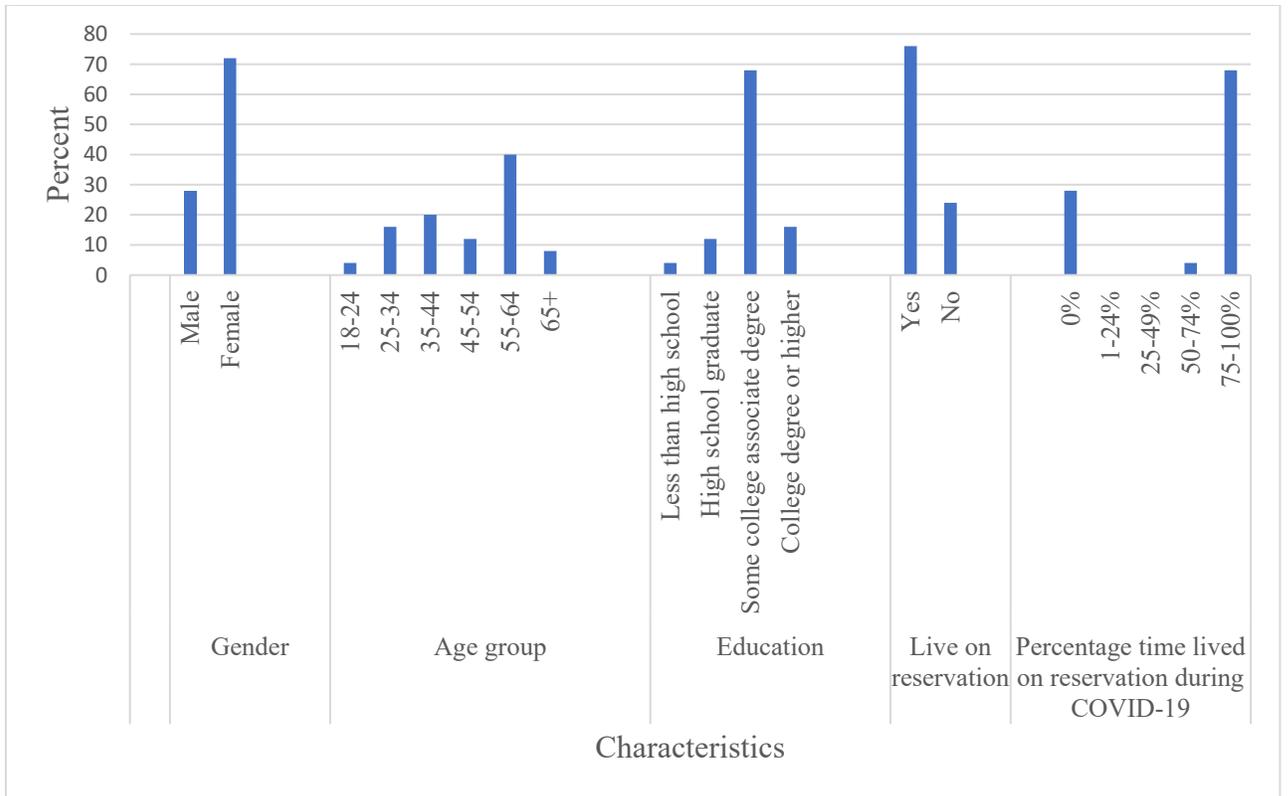

Figure 1. Demographic characteristics of the respondents

## 4.2 Preference for social media use

### 4.2.1 Preference for social media used for getting news and news headlines

Results presented in Figure 2 show that most of American Indians prefer Facebook (96%) followed by You Tube (60%), TikTok (36%), Instagram (24%), and Snapchat (16%) for getting news and news headlines. None of the respondents reported the use of Twitch, Nextdoor and WhatsApp for this purpose while a few (4%) reported the use of Reddit. LinkedIn, Twitter, and Pinterest were equally preferred (12%).

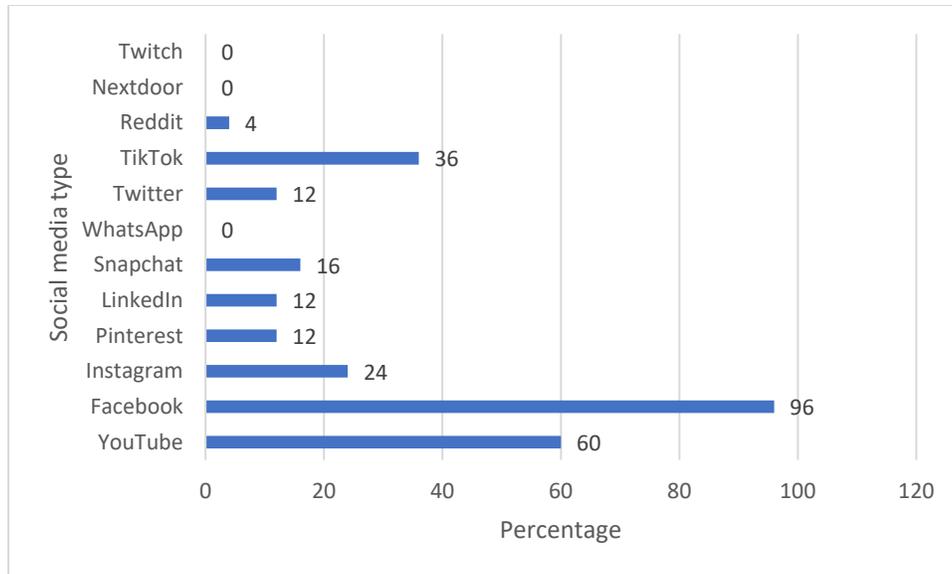

Figure 2. Use preference of social media for getting news and news headlines.

### 4.2.2 Frequency of social media use for news

Table 1 shows the summary of frequency of social media use for news by American Indians. Results show that Facebook, YouTube, and TikTok are the most frequently used social media type (with the highest number of respondents using it several times a day). The least frequently used social media types were WhatsApp, Twitter, Pinterest, Snapchat, and Instagram.

Table 1. Frequency of social media use for news

| Social media | 1=Several times a day | 2= Once a day | 3= Few times per week | 4= Once a week | 5= Every few weeks | 6= Less often | 7= No use |
|---|---|---|---|---|---|---|---|
| YouTube | 9 (36.00) | 1 (4.00) | 3 (12.00) | 0 (0.00) | 2 (8.00) | 2 (8.00) | 8 (32.00) |
| Facebook | 20 (80.00) | 4 (16.00) | | | | | 1 (4.00) |
| Instagram | 3 (12.00) | 3 (12.00) | 1 (4.00) | 1 (4.00) | | 1 (4.00) | 16 (64.00) |
| Pinterest | 3 (12.00) | 2 (8.00) | | | | | 20 (80.00) |
| LinkedIn | 1 (4.00) | | 2 (8.00) | | 1 (4.00) | 2 (8.00) | 7 (28.00) |
| Snapchat | 1 (4.00) | 1 (4.00) | 2 (8.00) | | 2 (8.00) | | 19 (76.00) |
| WhatsApp | | | | 1 (4.00) | | | 24 (96.00) |

| | | | | | | |
|---|---|---|---|---|---|---|
| Twitter | 1 (4.00) | | 1 (4.00) | 1 (4.00) | | | 22(88.00) |
| TikTok | 5 (20.00) | 1 (4.00) | 1 (4.00) | | 1 (4.00) | 2 (8.00) | 15(60.00) |
| Reddit | | | | | | 2 (8.00) | 23(92.00) |
| Nextdoor | | | | | 1 (4.00) | 1 (4.00) | 23(92.00) |
| Twitch | | | | | | | 25 100) |

### 4.3 Social media use and COVID-19

### 4.3.1 Impact of COVID-19 on social media use

During COVID-19 the use of social media increased tremendously. Most of the respondents (84%) reported that they increasingly used social media while only few (16%) reported the use was decreased (Figure 3). None of the respondents experienced any change in social media use. We have examined whether there is any difference in social media use between Native Americans who live on reservation versus those who live off-reservation, and we did not find any major differences.

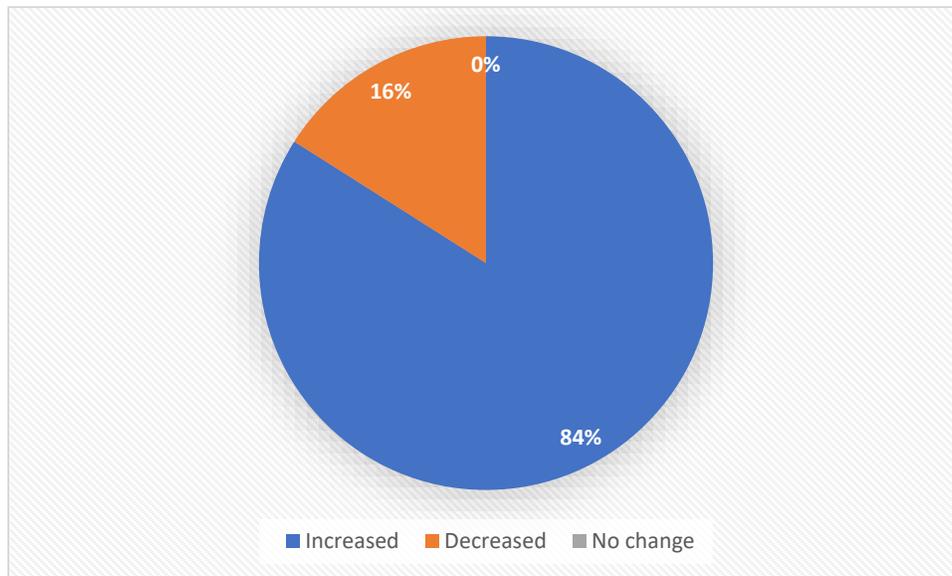

Figure 3. Change in social media use during COVID-19

### 4.3.2 Information received about COVID-19 and vaccines through social media use

Mostly used social media to get information about COVID-19 and vaccines were Facebook, YouTube, TikTok, and Instagram. None of the respondents used Nextdoor and Reddit for this purpose.

Table 2. Extent of use of social media for getting information regarding COVID-19 and vaccines

| Social media | 1= A lot | 2= Some | 3= Not much | 4= Not at all |
|---|---|---|---|---|
| YouTube | 6 (24.00) | 7 (28.00) | 4 (16.00) | 8 (32.00) |
| Facebook | 16 (64.00) | 5 (20.00) | 3 (12.00) | 4 (16.00) |
| Instagram | 3 (12.00) | | 3 (12.00) | 19 (76.00) |
| Pinterest | 1 (4.00) | 1 (4.00) | 1 (4.00) | 22 (88.00) |
| LinkedIn | 1(4.00) | | 2 (8.00) | 22 (88.00) |
| Snapchat | 1 (4.00) | 1 (4.00) | 2 (8.00) | 21 (84.00) |
| WhatsApp | | | 2 (8.00) | 23 (92.00) |
| Twitter | 2 (8.00) | | 2 (8.00) | 21 (84.00) |
| TikTok | 4 (16.00) | 1 (4.00) | 4 (16.00) | 16 (64.00) |
| Reddit | | | 1 (4.00) | 24 (96.00) |
| Nextdoor | | | | 25 (100.00) |
| Twitch | | | | 25 (100.00) |

### 4.4 Effect of social media use

### 4.4.1 Perceived effect of social media use

Most respondents (84%) reported that social media had mostly negative effects while few (28%) believed that it has mostly positive effects (Figure 4). Sixteen percent of the respondents had a neutral view on the effect of social media use.

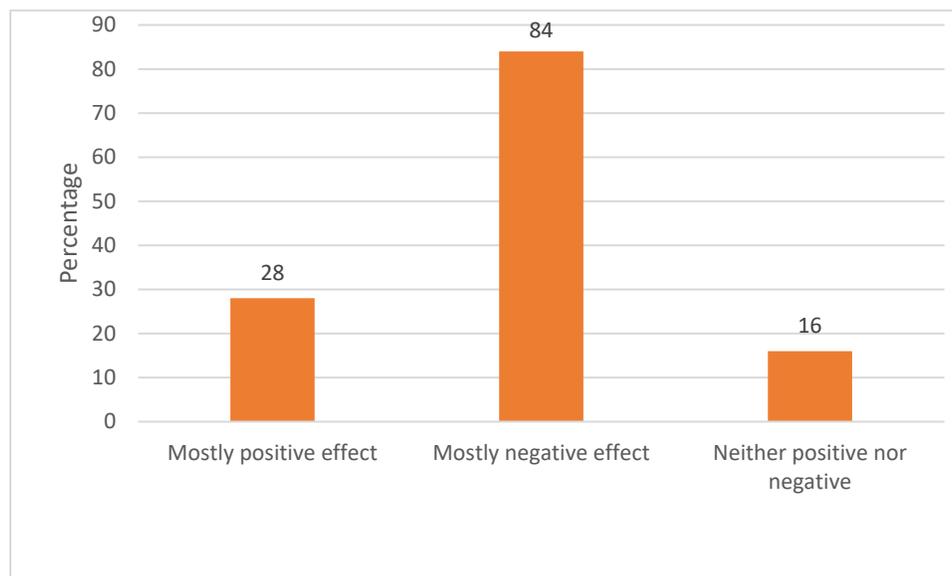
8

Figure 4. Perceived effect of social media use

### 4.4.2 Reasons cited for positive effects

The top three perceived positive effects include helping people stay informed and aware, communication/connection/community, and exposure to different options or viewpoints (Figure 5). The least preferred reason among the given alternatives was access to news.

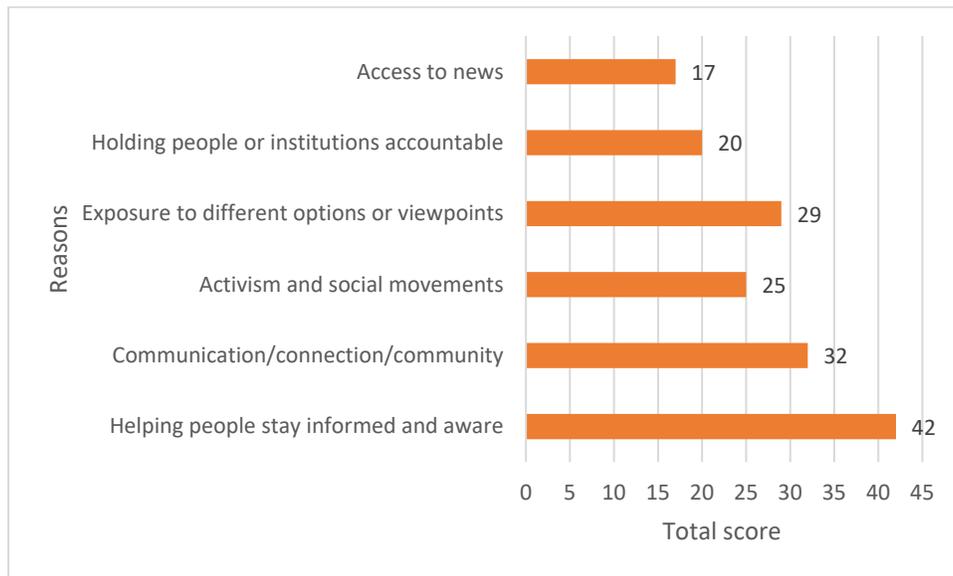

Figure 5. Reasons cited for positive effects

### 4.4.3 Reasons cited for negative effects

The top three reasons for negative effects of social media use include hate/harassment/extremism, misinformation/made up news, and people getting one point of view (Figure 6).

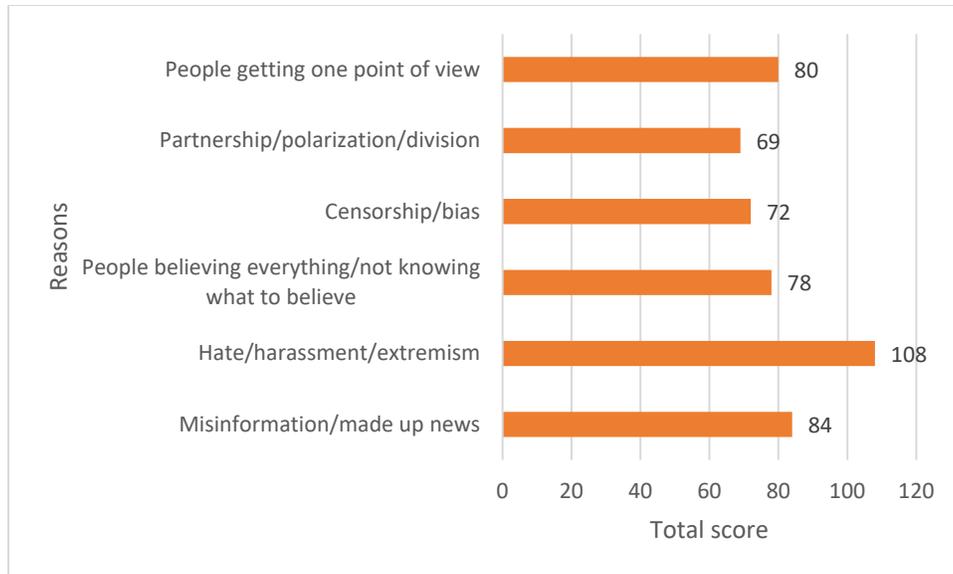

Figure 6. Reasons cited for positive effects

**Conclusions**

This study shows that the preferred social media platforms for American Indians in South Dakota are Facebook, YouTube, TikTok, Instagram and Snapchat. Their use of social media increased during the COVID-19 and they mainly used it for staying informed and aware, and for communication/connection within the community. Social media was an important source of information on COVID-vaccine and other COVID related information. Most of American Indians participated in the survey had negative perceptions about social media.

One of the limitations of the study is the small sample size. Because of this we were not able to look at whether social media use differs based on demographic characteristics such as age and sex. We did not find any difference in social media use between American Indians who live on reservations versus those who live off-reservation. We are not sure whether this is due to the small sample size of the study. Further studies are needed to ascertain whether there are any differences in social media use among American Indians living on reservations and off-reservations.

Zhao, Yuxin, Sixiang Cheng, Xiaoyan Yu, and Huilan Xu. 2020. "Chinese public's attention to the COVID-19 epidemic on social media: observational descriptive study." *Journal of medical Internet research* 22 (5):e18825.

# Appendix

**Table AI: Share daily experience through social media**

| Social media | 1=Several times a day | 2= Once a day | 3= Few times per week | 4= Once a week | 5= Every few weeks | 6= Less often | 7= No use |
|---|---|---|---|---|---|---|---|
| YouTube | 3(12.00) | 3(12.00) | | | | 3(12.00) | 16(64.00) |
| Facebook | 7(28.00) | 7(28.00) | 5(20.00) | | 3(12.00) | 2 (8.00) | 1 (4.00) |
| Instagram | 2 (8.00) | | | 1 (4.00) | 2 (8.00) | 2 (8.00) | 18(72.00) |
| Pinterest | 1 (4.00) | 1 (4.00) | 1 (4.00) | | 1 (4.00) | 2 (8.00) | 19(76.00) |
| LinkedIn | 1 (4.00) | | | | | 2 (8.00) | 22(88.00) |
| Snapchat | 1 (4.00) | 1 (4.00) | | 1 (4.00) | 1 (4.00) | 1 (4.00) | 20(80.00) |
| WhatsApp | 1 (4.00) | | | | | 1(4.00) | 23(92.00) |
| Twitter | | | | 2 (8.00) | 1 (4.00) | | 22(88.00) |
| TikTok | 2 (8.00) | 1 (4.00) | 1 (4.00) | | | 3(12.00) | 18(72.00) |
| Reddit | | | 1 (4.00) | | | | 24(96.00) |
| Nextdoor | | | | 1 (4.00) | | | 24(96.00) |
| Twitch | | | | 1 (4.00) | | | 24(96.00) |

**Table A2: Number of connected friends and/or family on social media**

| Social media | Mean | Median | Minimum | Maximum | Rank |
|---|---|---|---|---|---|
| YouTube | 8.52 | 3 | 0 | 50 | 5 |
| Facebook | 217.88 | 100 | 0 | 1000 | 2 |
| Instagram | 14.67 | 0 | 0 | 200 | 3 |
| Pinterest | 4.11 | 0 | 0 | 50 | 5 |
| LinkedIn | 3.05 | 0 | 0 | 50 | 5 |
| Snapchat | 404.4 | 0 | 0 | 7979 | 1 |
| WhatsApp | 0 | 0 | 0 | 0 | 9 |
| Twitter | 0.71 | 0 | 0 | 12 | 8 |
| TikTok | 13.42 | 2 | 0 | 140 | 4 |
| Reddit | 0 | 0 | 0 | 0 | 9 |
| Nextdoor | 0 | 0 | 0 | 0 | 9 |
| Twitch | 0 | 0 | 0 | 0 | 9 |

**Table A3: Frequency of reading posts in social media**

| Social media | 1=Several times a day | 2= Once a day | 3= Few times per week | 4= Once a week | 5= Every few weeks | 6= Less often | 7= Never |
|---|---|---|---|---|---|---|---|
| YouTube | 8 (32.00) | 4(16.00) | 5(20.00) | | | 3(12.00) | 5(20.00) |
| Facebook | 19(76.00) | 3(12.00) | 2 (8.00) | | | | 1 (4.00) |
| Instagram | 4 (16.00) | 2 (8.00) | 3(12.00) | 1 (4.00) | | 1 (4.00) | 14(56.00) |
| Pinterest | 2 (8.00) | 2 (8.00) | 1 (4.00) | | 2 (8.00) | 1 (4.00) | 17(68.00) |
| LinkedIn | 1 (4.00) | | | | | 3(12.00) | 21(84.00) |
| Snapchat | 2 (8.00) | 2 (8.00) | 1 (4.00) | | 1 (4.00) | 2 (8.00) | 17(68.00) |
| WhatsApp | | | | | | | 25(100.0) |
| Twitter | 1 (4.00) | 1 (4.00) | | | | | 23(92.00) |
| TikTok | 5 (20.00) | 1 (4.00) | 1 (4.00) | 1 (4.00) | 1 (4.00) | 4 | 12(48.00) |
| Reddit | | 1 (4.00) | | | | 1 (4.00) | 23(92.00) |
| Nextdoor | | | | | | 1 (4.00) | 24(96.00) |
| Twitch | | | | | | 1 (4.00) | 24(96.00) |

**Table A4: Frequency of posting in social media**

| Social media | 1=Several times a day | 2= Once a day | 3= Few times per week | 4= Once a week | 5= Every few weeks | 6= Less often | 7= Never |
|---|---|---|---|---|---|---|---|
| YouTube | 5 (20.00) | 1 (4.00) | 1 (4.00) | | | 4(16.00) | 14(56.00) |
| Facebook | | | | | | | |
| Instagram | 2 (8.00) | | 1 (4.00) | 1 (4.00) | 1 (4.00) | 3(12.00) | 17(68.00) |
| Pinterest | 1 (4.00) | 2 (8.00) | | | 1 (4.00) | 2 (8.00) | 19(76.00) |
| LinkedIn | 2 (8.00) | | | | | 4(16.00) | 19(76.00) |
| Snapchat | 3(12.00) | | 1 (4.00) | | 1 (4.00) | 2 (8.00) | 18(72.00) |
| WhatsApp | 1 (4.00) | | | | | 2 (8.00) | 22(88.00) |
| Twitter | 2 (8.00) | | 1 (4.00) | | | | 22(88.00) |
| TikTok | 2 (8.00) | 3(12.00) | 1 (4.00) | | | 1 (4.00) | 18(72.00) |
| Reddit | 2 (8.00) | | | | | | 23(92.00) |
| Nextdoor | 1 (4.00) | | | | | | 24(96.00) |
| Twitch | | | 1 (4.00) | | | | 24(96.00) |

**Table A5: Post comments on social media**

| Post comments | Frequency | Percent |
|---|---|---|
| Never (1) | 13 | 52.00 |
| About once every year (2) | 0 | 0.00 |
| About once every half year (3) | 1 | 4.00 |
| About once every 3 months (4) | 3 | 12.00 |
| About once every month (5) | 3 | 12.00 |
| About once every week (6) | | |
| About once every day (7) | 3 | 12.00 |
| More than once per day (8) | 2 | 8.00 |
| Total | 25 | 100.0 |

**Table A6: Number of users followed by respondents**

| Social media | Mean | Median | Minimum | Maximum |
|---|---|---|---|---|
| YouTube | 56.14 | 0 | 0 | 1000 |
| Facebook | 138.95 | 25 | 0 | 2000 |
| Instagram | 108.59 | 0 | 0 | 2000 |
| Pinterest | 91.77 | 0 | 0 | 2000 |
| LinkedIn | 91.22 | 0 | 0 | 2000 |
| Snapchat | 95.45 | 0 | 0 | 2000 |
| WhatsApp | 0 | 0 | 0 | 0 |
| Twitter | 92.36 | 0 | 0 | 2000 |
| TikTok | 215.5 | 0 | 0 | 2000 |
| Reddit | 0 | 0 | 0 | 0 |
| Nextdoor | 0 | 0 | 0 | 0 |
| Twitch | 0 | 0 | 0 | 0 |

**Table A7: Reliable source of media**

| Source of media/ Scale | 1=Not reliable | 2= Slightly reliable | 3= Somewhat reliable | 4= Quite reliable | 5= Very reliable | Total score | Rank |
|---|---|---|---|---|---|---|---|
| Newspaper | 1 | 1 | 4 | 12 | 3 | 78 | 1 |
| Public radio | 1 | 2 | 4 | 10 | 4 | 77 | 2 |
| Website | 1 | 4 | 6 | 9 | 1 | 68 | 3 |
| Social media | 6 | 6 | 6 | 3 |  | 48 | 4 |

*Note: Total score is calculated by summing up all the products of scale and frequency of each social media type. For example, $Total\ score\ for\ newspaper = 1 \times 1 + 2 \times 1 + 3 \times 4 + 4 \times 12 + 5 \times 3 = 78$*

**Table A8: Reliability of different social media**

| Source of media/Scale | 1=Not reliable | 2= Slightly reliable | 3= Somewhat reliable | 4= Quite reliable | 5= Very reliable | Total score | Rank |
|---|---|---|---|---|---|---|---|
| YouTube | 3 | 1 | 6 | 2 | 3 | 46 | 1 |
| Facebook | 1 | 3 | 8 | 1 | 2 | 45 | 2 |
| Instagram | 5 | 4 | 4 |  | 2 | 35 | 3 |
| Pinterest | 6 | 5 | 3 |  | 1 | 30 | 5 |
| LinkedIn | 6 | 3 | 3 | 1 | 2 | 35 | 3 |
| Snapchat | 8 | 3 | 2 |  | 2 | 30 | 5 |
| WhatsApp | 9 | 2 | 3 |  | 1 | 27 | 9 |
| Twitter | 6 | 5 | 3 |  | 1 | 30 | 5 |
| Tiktok | 8 | 2 | 4 |  | 1 | 29 | 8 |
| Reddit | 9 | 3 | 2 |  | 1 | 26 | 10 |
| Nextdoor | 11 | 1 | 1 | 1 | 1 | 25 | 11 |
| Twitch | 10 | 2 | 2 |  | 1 | 25 | 11 |

*Note: Total score is calculated by summing up all the products of scale and frequency of each social media. For example, $Total\ score\ for\ YouTube = 1 \times 3 + 2 \times 1 + 3 \times 6 + 4 \times 2 + 5 \times 3 = 46$*